\documentclass[%
 reprint,
 amsmath,amssymb,
 aps,prb
]{revtex4-2}

\usepackage{xcolor} 
\usepackage{graphicx}
\usepackage{dcolumn}
\usepackage{bm}

\DeclareMathOperator{\e}{e}

\newcommand{\vect}[1]{{\bf{#1}}}

\begin{document}

\title{Spin relaxation in diluted magnetic semiconductors. GaMnAs as example.}


\author{I.V. Krainov}
\email{igor.kraynov@mail.ru} 
\author{V.F. Sapega}
\email{sapega.dnm@mail.ioffe.ru} 
\author{G.S. Dimitriev}
\author{N.S. Averkiev}
\affiliation{Ioffe Institute, 194021 St. Petersburg, Russia}

\date{\today}

\begin{abstract}
We report on study of magnetic impurities spin relaxation in diluted magnetic semiconductors above Curie temperature. Systems with a high concentration of magnetic impurities where magnetic correlations take place were studied. The developed theory assumes that main channel of spin relaxation is mobile carriers providing indirect interactions between magnetic impurities. Our theoretical model  is supported by experimental measurements of manganese spin relaxation time in GaMnAs by means of spin-flip Raman  scattering. It is found that with temperature increase spin relaxation rate of ferromagnetic samples increases and tends to that measured in paramagnetic sample.
\end{abstract}

\maketitle

\section{Introduction}

Diluted magnetic semiconductors (DMS) are promising system as functional materials combining semiconductor and magnetic properties \cite{SpinRevModPhys, RevModPhys.86.187}. 
In order to construct, manipulate and incorporate these materials in conventional electronics it is necessary to understand the mechanisms of formation and destruction of magnetic order. At present,  mechanism of magnetic ordering is well studied \cite{Jungwirth}, but the processes of spin relaxation in these correlated systems have not been fully understood. This paper will address this question. 

The nature of ferromagnetism in these materials is related to exchange interaction between localized spins and carriers \cite{KittelOriginal,Dietl1019,Dietl2001}. The ferromagnetic interaction between magnetic moments of impurities can be mediated by band electrons or holes like in CdMnTe or localized carriers forming impurity band like GaMnAs. In GaAs doped with Mn, the $S = 5/2$ local moments of the Mn ions are exchange coupled with holes provided by substituting of Ga with Mn  \cite{PRBsingleMn,SapegaFTT,MnGPRL,AverkievFTT, Linnarsson1997}. In low doping regime, when average distance between ions exceeds Mn acceptor radius,  exchange interaction between manganese $d$ shell electrons with total spin $S = 5/2$ and hole with angular momentum $J = 3/2$ results in the splitting of ground state into four levels with total angular momentum $F=1 \div 4$ \cite{SapegaPSSb,PhysRevB.93.235202}.  

An increase in the Mn concentration leads to overlapping of acceptor bound holes \cite{Richardella665}, which causes diffusion of spins and charges. After this happened the indirect exchange interaction between manganese via holes begins to determine main magnetic properties in the  system. Depending on ratio between average value of indirect exchange and its dispersion the spin glass or ferromagnetic ordering takes place at low temperatures \cite{spinGlassRev}. 
At high temperatures ($T>T_C$) there is no magnetic order and subsystem of magnetic ions can be considered as many body interacting system with correlations. Another system demonstrating very similar properties is ideal crystal with free band carriers doped with magnetic impurities. For high concentration of magnetic impurities the spin glass or ferromagnetic ordering takes place when RKKY (Ruderman–Kittel–Kasuya–Yosida) temperature higher than Kondo temperature. In paramagnetic phase we have also correlated spin system consisting of magnetic centers and delocalized carriers. Thus in both these systems the correlation between magnetic ions is mediated by mobile carriers. Besides, these carriers provide the main channel of magnetic impurities spin relaxation. In this paper we analyze spin relaxation of magnetic localized spins and describe its different regimes in paramagnetic phase. Our theory is supported by experimental measurements of manganese transversal spin relaxation rate in GaMnAs with Mn concentration up to few percent including both paramagnetic samples (with Curie temperature lower than that liquid helium) and ferromagnetic samples with Curie temperatures of dozens of kelvins. The theoretical model is compared with data obtained using spin flip Raman scattering (SFRS) technique, which as has been shown recently (see Ref.\cite{OURPRBGaMnAs}) allows directly to measure manganese spin  
relaxation time. For spin relaxation in ferromagnetic samples, two temperature-dependent mechanisms are determined.

\section{Experiment}

\begin{figure*}
	\centering
	\includegraphics[width=0.49\linewidth]{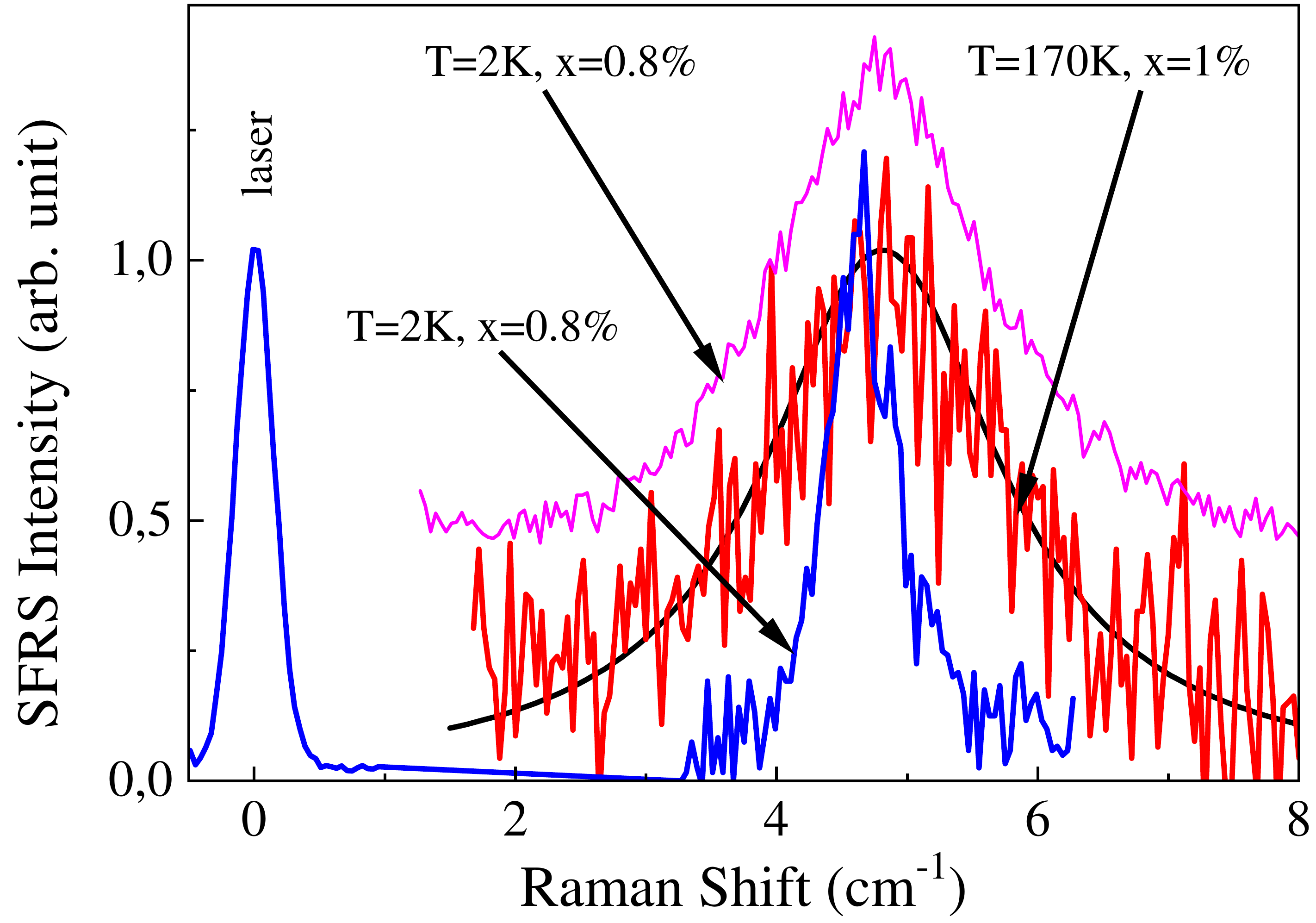}
	\includegraphics[width=0.43\linewidth]{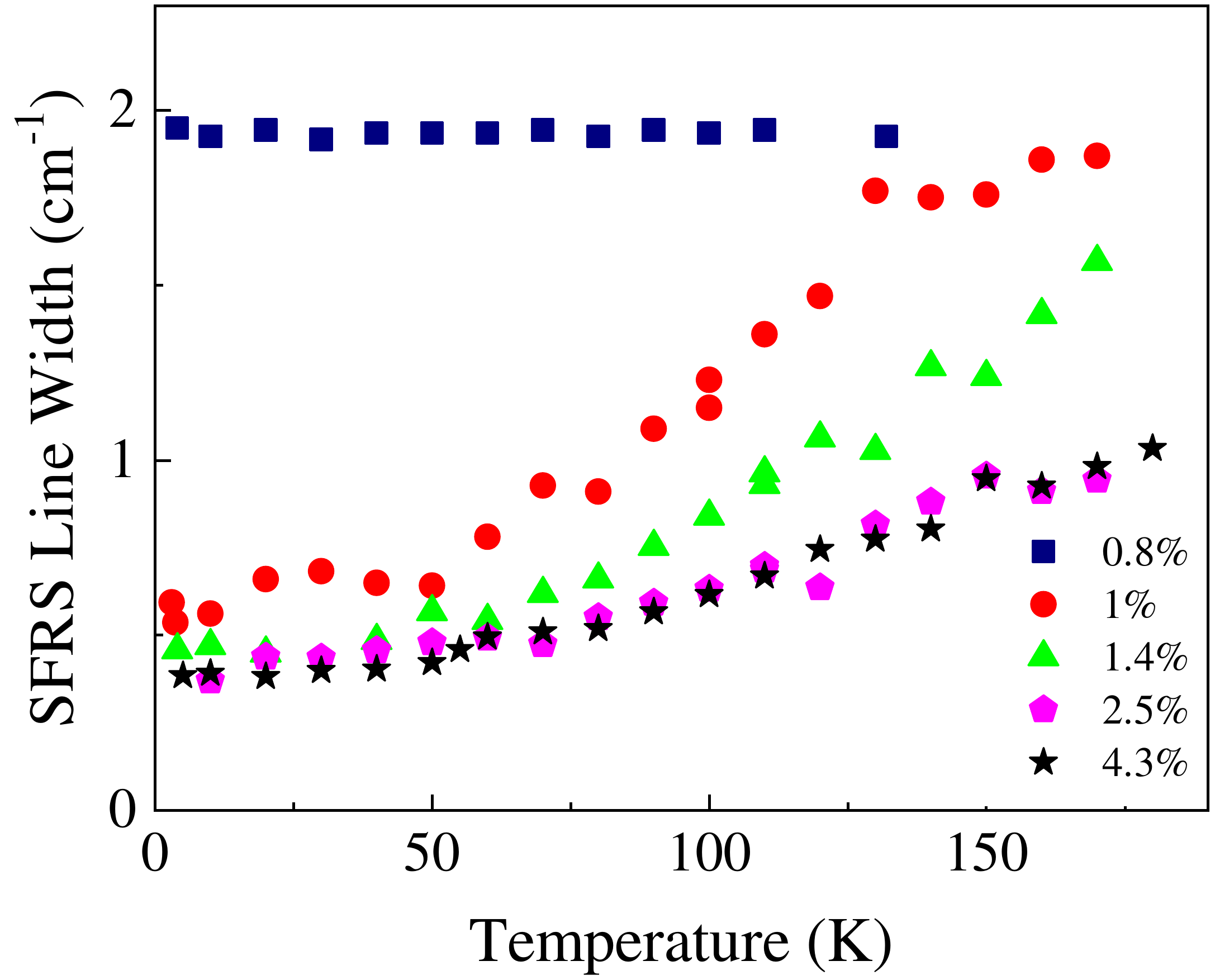}
	\caption{(Left panel) Spin flip Raman scattering spectra measured
in GaMnAs samples with a concentration of manganese x = 0.8 \%, 1 \%. The measurements were carried out in an external magnetic field $B = 5$~T and at different temperatures. (Right panel) Temperature dependence of SFRS line width (proportional to manganese spin relaxation rate) for samples with different Mn concentration, from x = 0.8 \% up to x = 4.3 \% measured in an  external magnetic field $B=5$~T.}
	\label{fig}
\end{figure*}

The samples for the present study were grown by molecular-beam
epitaxy on GaAs (001) substrates (semiinsulating, Si, or Zn doped). The samples have a Mn content of
$x\sim 0.008\div0.07$. 

The SFRS spectra were analyzed by a Jobin-Yvon U1000 double
monochromator, equipped with a cooled GaAs photomultiplier and
conventional photon counting electronics. For the excitation of
SFRS, we used the lines of He-Ne, Kr-ion and tunable Ti-saphier lasers. The laser power densities focused on the sample was about 50 Wcm$^{-2}$. The
experiments were carried out in the temperature range 1.6–200 K and
magnetic fields up to 5 T in backscattering Voigt geometry. In the
Voigt geometry Raman spectra were recorded in $x(\sigma,\pi)\bar{x}$
polarization configuration, with $\bar{x}$ and $x$ perpendicular to
the sample plane and the magnetic field $B$ directed along $z$ and
$\sigma$, $\pi$,  denoting linear polarizations of the exciting
($\sigma$) and scattered ($\pi$) light with the electric field
vector of the light perpendicular (for $\sigma$) or parallel (for
$\pi$) to $B$.

Raman spectra (see left panel of Fig.\ref{fig}) of two diluted (Ga,Mn)As samples: 
paramagnetic PM ($x = 0.8\%$, solid magenta  line, $T_C < 2$~K) and FM1 ($x = 1\%$, solid blue line, $T_C \approx 35$~K) recorded
in a magnetic field $B=5$~T at temperatures $T=2$~K show a strongly
polarized Mn-SFRS line with a magnetic - field - dependent Raman
shift. These SFRS lines are active only in $x(\sigma,\pi)\bar{x}$ or
$x(\pi,\sigma)\bar{x}$ configuration (with $\sigma
\parallel [001]$) and forbidden in $x(\sigma,\sigma)\bar{x}$ and
$x(\pi,\pi)\bar{x}$. The magnetic-field dependence of the Mn-SFRS
line energy shift in PM sample can be represented as $\Delta = g_0^
{Mn}\mu_BB$ with $g_0^{Mn} = 2.01$, which does not depend on
temperature. In contrast in the FM samples the magnetic field
dependence of the Raman shift for the SFRS line can be represented
as $\Delta = g_0^{Mn}\mu_BB$ with $g_0^{Mn} = 2.01$ only at $T$ above
Curie temperature $T_C$. Below $T_C$ g-factor of the FM sample
depends on the temperature $g_0^{Mn} = g(T)$.\cite{OURPRBGaMnAs}. The SFRS
linewidth in the PM and FM samples does not depend on magnetic
field. It means that the measured Mn SFRS linewidth is determined
exclusively by the transversal lifetime of the Mn spin
subsystem. Therefore the Mn spin subsystem lifetime can be
determined from the SFRS linewidth. In PM sample the SFRS linewidth
is not sensitive to temperature in wide range of temperatures as
it is shown by deep blue squares in right panel of Fig.~\ref{fig}. In contrast the SFRS line
width of the FM sample strongly depends on temperature as one can
see from comparison of Raman spectra measured at $T=2$~K (solid blue
line in left panel of Fig.\ref{fig}) and $T=170$~K (solid  red line). The
temperature dependence of the SFRS linewidth measured at $B = 5$~T for
a few FM samples is shown in right panel of Fig.\ref{fig}. In FM1 sample the
SFRS linewidth is not sensitive to temperature below $T_C$ and
strongly increases at $T>T_C$. Further increase of temperature
($T>130$~K) leads to linewidth saturation at a value close to PM
sample line width. The linewidth of FM samples with higher Mn
content demonstrate similar to the FM1 sample behavior. However the Mn
content increase reveals itself in decrease of the slope of the
temperature dependence of line width at $T>T_C$ and absence of
saturation in high temperature range. Note that the SFRS signal
decreases strongly in all samples at $T>150\div180$~K what does not
allow to reach saturation in samples with Mn content higher than
1.4\%.


\section{Theory}

The structure under investigatigation is bulk semiconductor doped with magnetic impurities. If their is some delocalized carriers the indirect exchange (RKKY) between magnetic impurities takes place \cite{KittelOriginal,Dietl2001}. The delocalized carriers can appear by impurity band formation (for example if magnetic impurities are donors or acceptors, like GaMnAs) or it could be band electrons/holes. In general, developed theory can also be valid for localized carriers, but it is necessary that carrier's wave function overlap with many magnetic impurities. With increasing concentration of magnetic impurities indirect exchange interaction will lead to ferromagnetic transition at low temperatures \cite{Jungwirth}.  

We will describe the magnetic properties using two magnetic subsystems with their own magnetizations \cite{KittelG, OURPRBGaMnAs}. 
One consists of magnetic impurities spins $M_S(x,t)$ and second is of delocalized carriers spins $M_J(x,t)$. These two subsystems interact with each other via exchange interaction with constant $\lambda$. This approach was successfully applied to study ferromagnetic resonance in rare earth garnets \cite{KittelG}. We will analyze magnetization dynamics of these  two subsystems to obtain magnetic impurity tranversal spin relaxation time \cite{OURPRBGaMnAs}. For GaMnAs these subsystems are Mn spins and holes spins.

We focus on paramagnetic region ($T > T_C$, $T_C$ - Curie temperature) where the homogeneous magnetic field induced  magnetization is small. In order to find transversal spin relaxation one needs to find homogeneous modes frequency of magnetization precession in magnetic field. Our theory will be phenomenologic, we want to take into account only main ingredients. Therefore we write the most simple equations of magnetization motion in external magnetic filed:
\begin{gather}
	\frac{d \vect{M}_S}{d t} =  \frac{\mu_B g_S}{\hbar} \vect{B} \times \vect{M}_S - 
		\lambda \frac{\mu_B g_S}{\hbar} \vect{M}_J \times \vect{M}_S, \\ \nonumber
	\frac{d \vect{M}_J}{d t} =  \frac{\mu_B g_J}{\hbar} \vect{B} \times \vect{M}_J - 
		\lambda \frac{\mu_B g_J}{\hbar} \vect{M}_S \times \vect{M}_J + \\ 
		+ D \triangle \vect{M}_J - \gamma_J (\vect{M}_J - \vect{M}^0_J).
\end{gather}
where $g_S$, $g_J$ - are g-values of magnetic ions and delocalized carriers, $D$, $\gamma_J$ - are spin diffusion coefficient and relaxation rate of delocalized carriers, $\vect{M}^0_J$ - is homogeneous magnetic field induced magnetization, $B$ - external magnetic field applied along $z$ axis. These equations may be supplemented with magnetic anisotropy field, shape field, but for a large applied field they could be neglected. 
We assume that magnetic impurities mainly interact with delocalized carriers and neglect other contributions like spin-phonon, spin-nuclear interactions and so on. 
This assumption looks reasonable because only mobile carriers can mediate an interaction between distant localized spins, and simultaneously to serve as a spin relaxation channel. Spin diffusion is governed by carrier diffusion and carrier spin-spin interaction. The spin relaxation is caused by spin-orbit and spin-spin interactions. Both of this parameters can depend on temperature, carriers and magnetic impurities concentrations. The magnetization of magnetic impurities subsystem can be presented in following form $\vect{M}_S = \vect{M}_S^0 + \vect{M}_S(t)$ where     
$\vect{M}_S^0$ is homogeneous magnetic field induced magnetization        
and $\vect{M}_S(t)$ is small homogeneous deviation of magnetization       
precessing in external field. For delocalized carriers, the subsystem magnetization has an additional fluctuating term $\vect{M}_J = \vect{M}_J^0 + \vect{M}_J(t) + \delta \vect{M}_J(x,t)$ due to the fact that they are effectively coupled with phonons.

In paramagnetic phase magnetic field induced magnetization is small in comparison with external magnetic field $\lambda M^0_i \ll B$. Otherwise the presence of $ M ^ 0_i $ leads to a renormalization of the g-factor \cite{KittelG}, as it was observed in GaMnAs in the ferromagnetic phase \cite{OURPRBGaMnAs}.  
Let us find the frequency of the homogeneous mode under the assumption that its amplitude is infinitely small, i.e. less than fluctuation $\delta M_J(x,t) \gg M_{S,J}(t)$.  

We introduce the following variables to simplify the equations of motion of the magnetization
$\mathcal{B}_J = \vect{B} \mu_B g_J / \hbar$, $\mathcal{B}_S = \vect{B} \mu_B g_S / \hbar$, $\lambda_S = \lambda \mu_B g_S / \hbar$, $\lambda_J = \lambda \mu_B g_J / \hbar$, $M^\pm = M^x \pm i M^y$.
Under these assumptions, the equation for the fluctuating part of the hole magnetization is:
\begin{gather}
\label{dMj}
\frac{d \delta \vect{M}_J}{d t} =  \vect{\mathcal{B}}_J \times \delta \vect{M}_J + D \triangle \delta \vect{M}_J - \gamma_J \delta \vect{M}_J,
\end{gather}
and for homogeneous parts of magnetization:
\begin{gather}
\label{eqMjz}
\frac{d M^+_J}{d t} = - i \mathcal{B}_J M^+_J + i \lambda_J [ (M^+_J + \delta M^+_J) M^z_S - M^z_J M^+_S ] - \\ \nonumber 
- \gamma_J \delta M^+_J, \\ \label{eqMjp}
\frac{d M^z_J}{d t} = \frac{i}{2} \lambda_J [ (M^-_J + \delta M^-_J) M^+_S - (M^+_J + \delta M^+_J) M^-_S ], \\ \label{eqMsp}
\frac{d M^+_S}{d t} = - i \mathcal{B}_S M^+_S + i \lambda_S [ M^z_J M^+_S - (M^+_J + \delta M^+_J) M^z_S ], \\ \label{eqMsz}
\frac{d M^z_S}{d t} = \frac{i}{2} \lambda_S [ (M^+_J + \delta M^+_J) M^-_S - (M^-_J + \delta M^-_J) M^+_S ].
\end{gather}
There are two scenarios for the spin relaxation of the magnetic impurity depending on the rate of its rotation in the characteristic fluctuation field $\lambda_S \delta M_J$ and the time of fluctuations dissipation $\tau_{dis}$. For the case of strong fluctuations or their slow dynamics $\lambda_S \delta M_J \tau_{dis} \gg 1$ the spin relaxation time of magnetic impurities is determined by the dissipation time of $\gamma_S \approx 1/\tau_{dis}$. We will calculate this equation later.

Otherwise $\lambda_S \delta M_J \tau_{dis} \ll 1$ it is necessary to take into account the dynamics of fluctuations. To do this, we integrate the equations (\ref{eqMsz}, \ref{eqMjz}), then substitute them into the equation (\ref{eqMsp}) and average over the fluctuations.
After that we get: 
\begin{gather}
\nonumber 
	\frac{d M^+_S(t)}{d t} = - i \mathcal{B}_S M^+_S (t) - \\ \label{eqMs}
	- \frac{\lambda^2_S}{2} 
		\int^t d\tau\, \langle \delta M^+_J (x,t) \delta M^-_J (x,\tau) \rangle  M^+_S (\tau).
\end{gather}
From here we can obtain the equation for the frequency $\omega_{S} = \mathcal{B}_S - i I(\omega_{S})$, where $I(\omega)$ is Fourier transform of last part of (\ref{eqMs}). Assuming that the spin are precess $\mathcal{B}_S > I(\mathcal{B}_S)$ we have for spin relaxation rate:
\begin{gather}
\gamma_{S} \approx \operatorname{Re} \left\{ \lambda^2_S \int \frac{d\vect{k}}{(2 \pi)^3}  G(k, \mathcal{B}_S) K(k) \right\},
\end{gather}
where $K(x) = \langle \delta M^x_J (x) \delta M^x_J (0) \rangle$ - is the static fluctuation correlator, $G(k, \omega) = [ Dk^2 + \gamma_J - i (\omega - \mathcal{B}_J) ]^{-1}$ is Green function of transverse components of the Equation (\ref{dMj}), $\operatorname{Re}$ - being the real part.

The correlator of static fluctuations can be obtained from the phenomenological Hamiltonian by expanding it in a power series in  $\delta M_J$ to quadratic terms:
\begin{gather}
\label{Heff}
H_{eff}(\delta M_J) = C \frac{\delta M_J^i}{\partial x_j} \frac{\delta M_J^i}{\partial x_j} + \alpha \delta M_J^i\delta M_J^i,
\end{gather}
where $ C $ is the square of the characteristic scale of magnetic correlations and, as it is known, this constant is proportional to the Curie temperature ($C \sim T_C$). Constant $\alpha$ defines the magnetic susceptibility. 
One has for correlator:
\begin{gather}
\nonumber
K(x) = \frac{ \int D[\delta M_J] \e^{ - \int dx\, H_{eff}(\delta M_J)/T } \delta M^x_J (x) \delta M^x_J (0) }{ \int D[\delta M_J] \e^{ -\int dx\,  H_{eff}(\delta M_J)/T } }, \\ \label{Kk}
K(k) = \frac{T} {2(Ck^2+\alpha)}.
\end{gather}
After this we can obtain magnetic impurities spin relaxation rate:
\begin{gather}
\gamma_{S} = \operatorname{Re} \left\{
\frac{\lambda^2_S}{8 \pi} \frac{T}{D \sqrt{C \alpha} + C \sqrt{D [ \gamma_J - i (\mathcal{B}_S - \mathcal{B}_J)  ]} } \right\}. 
\end{gather}
The asymptotic for high magnetic field $|\mathcal{B}_S - \mathcal{B}_J| \gg D \alpha / C$ is $\gamma_{S} = \lambda^2_S T / 8\pi C\sqrt{2 D |\mathcal{B}_S - \mathcal{B}_J|}$. 
For coefficients almost independent of temperature in this equation, it predicts a linear increase in the spin relaxation rate with temperature. This tendency agrees well with the linear increase in the relaxation rate in the paramagnetic phase with temperature observed in GaMnAs.
But such a linear growth looks unrealistic, especially at high temperatures. This also contradicts the observed saturation of the relaxation time of the Mn spin versus temperature for the GaMnAs samples, see right panel of  Fig.\ref{fig}.
The fluctuation correlator (\ref{Kk}) has radius $r_c = \sqrt{C / \alpha}$ and fluctuation power $\int dx\, K(x) = T / 2 \alpha$ which grows infinitely with temperature. To get it one needs to have more and more delocalized carriers in the region with the scale $ r_c $. But the process stops when it reaches a certain maximum value of the fluctuation. This is limited by the Coulomb interaction between carriers.
The maximum fluctuation is one full spin of the delocalized carrier at its Fermi wavelength. For example, for GaMnAs, this is the total spin of one hole on one manganese. To take this into account, we need to know the fluctuation Hamiltonian (\ref{Heff}) for high values of $\delta M_J$ in  nonlinear regime. We can phenomenologically take into account the saturation effect by introducing a cutoff into the integration over the fluctuating field:
\begin{gather}
\e^{ -\int dx\, H_{eff}(\delta M_J)/T } \rightarrow \e^{ -\int dx\, H_{eff}(\delta M_J)/T - \int dx\, (\delta M_J)^2/M^2_0}.
\end{gather}
The value $M_0$ limits fluctuations and is the maximum fluctuation of the spin of delocalized carriers, which can be on any magnetic impurity. It  works like:
\begin{gather} 
\nonumber
K(r) = \frac{1}{8\pi} \frac{T}{C} \frac{\e^{-r \sqrt{\alpha/C}}}{r}, \quad M_0 \rightarrow \infty, \\ \nonumber 
K(r) = M_0^2 \delta(r), \quad M_0 \rightarrow 0.
\end{gather}
For any temperature the correlator will be:
\begin{gather}
\label{corrEx}
K(k) = \frac{T} {2(Ck^2+\alpha + T / M_0^2)}.
\end{gather}
Using this we obtain general equation of magnetic impurities spin relaxation rate:
\begin{widetext}
\begin{gather}
\nonumber
	\gamma_{S} = \frac{\lambda^2_S}{4\pi^2}  \operatorname{Re} \bigg\{ 
	\int_0^{1/a} dk\, k^2 \frac{1}{D k^2 + \gamma_J - i (\mathcal{B}_S - \mathcal{B}_J)}  \frac{T} {Ck^2+\alpha + T / M_0^2} \bigg\} = \\ 
	= \frac{\lambda^2_S}{4\pi^2}  \operatorname{Re} \biggl\{ \frac{T} {\gamma_J l_D C} 
	\frac{ \sqrt{\frac{\alpha M_0^2 + T}{C M_0^2 l_D^{-2}}} \arctan \sqrt{\frac{C M_0^2 a^{-2}}{\alpha M_0^2 + T}} - \sqrt{1 - i (\mathcal{B}_S - \mathcal{B}_J) / \gamma_J} \arctan \frac{l_D}{a \sqrt{1 - i (\mathcal{B}_S - \mathcal{B}_J) / \gamma_J}}
	}{
	\frac{\alpha M_0^2 + T}{C M_0^2 l_D^{-2}} - 1 + i (\mathcal{B}_S - \mathcal{B}_J) / \gamma_J
	} \biggl\},
	\label{gMnInt}
\end{gather}
\end{widetext}
where $l_D = \sqrt{D / \gamma_J}$ is the spin diffusion length. Here we have limited the integral over the Fermi wavelength of delocalized carriers $a$.

This equation describes various scenarios of spin relaxation of magnetic impurities in the DMS for the paramagnetic phase. For ferromagnetic samples (with a Curie temperature higher than the temperature of liquid helium), it has a linear increase in temperature (provided that all coefficients are close to constant in temperature), and then goes into saturation due to the attainment of the maximum fluctuation. The maximum fluctuation is determined by one spin of the delocalized carrier located at the Fermi wavelength $\delta M_J^{max} \sim J \mu_B / a^3 \sim M_0 a^{-3/2}$. The saturation temperature defined by minimum of $T_1 = M_0^2 C a^{-2}$ and $T_2 = M_0^2 \alpha$. It is important to clarify how this temperature depends on the carrier concentration. The spin correlation scale $ C \sim T_C $ is related with the Curie temperature. Thus, for systems with a low Curie temperature, the spin relaxation rate of magnetic impurities is in saturation.

For systems which are paramagnetic at all temperatures (with Curie temperature lower than liquid helium), or for ferromagnetic samples in saturation region (Eq.\ref{gMnInt} for high temperatures) the fluctuations are strong because they reach their maximum value:   
\begin{gather} 
\label{gMnSat1}
\gamma_{S} \approx
\frac{\lambda^2_S}{4\pi^2} \frac{M_0^2}{D a}, \quad 
D a^2 \gg \gamma_J, |\mathcal{B}_S - \mathcal{B}_J|, \\
\gamma_{S} \approx \frac{\lambda^2_S}{12\pi^2} \frac{M_0^2 \gamma_J}{a^3 (\gamma_J^2 + |\mathcal{B}_S - \mathcal{B}_J|^2)}, \quad 
D a^2 \ll \gamma_J, |\mathcal{B}_S - \mathcal{B}_J|.
\label{gMnSat2}
\end{gather}
This equation is quite natural and describes the fluctuation field of the spin $ J $ acting on the spin of magnetic impurities, and the fluctuations dissipation time is the diffusion spreading of the $ a ^ 3 $ region or spin relaxation.
In the case of strong fluctuations or their slow dynamics $ \delta M_J \tau_ {dis} \gg 1 $, as we mentioned earlier, the spin relaxation time of magnetic impurities is determined by fluctuations dissipation time $\gamma_S \approx 1/\tau_{dis}$. This time is the minimum of the relaxation time of the spin of delocalized carriers or the diffusion propagation of fluctuations $\gamma_S \approx \min[\gamma_J, D/\max(C, a^2)]$. Note that in the case of slow dynamics, some other spin relaxation processes such as hyperfine interaction may be important.

\section{Discussion and comparison with experiment}

For the GaMnAs with concentrations about atomic percent estimates show that $ \delta M_J \tau_ {dis} \ll 1 $ and one needs to use equation~(\ref{gMnInt}) for manganese spin relaxation rate. 
The cutoff length $a$ in this DMS is determined by localized hole radius. The maximum fluctuation is one hole spin on one manganese ion. For paramegnetic samples with $x < 1 \%$, the spin relaxation rate independent on temperature right panel of Fig.~\ref{fig}, which predicted by eq.~(\ref{gMnSat1}, \ref{gMnSat2}). This saturation is related to low Curie temperature in these samples. For samples with $ x> 1 \% $, which have a Curie temperature of several tens of Kelvin, the spin relaxation rate increases linearly with temperature in the paramagnetic region and then tends to saturation. 

\begin{figure}[bh]
	\centering
	\includegraphics[width=0.95\linewidth]{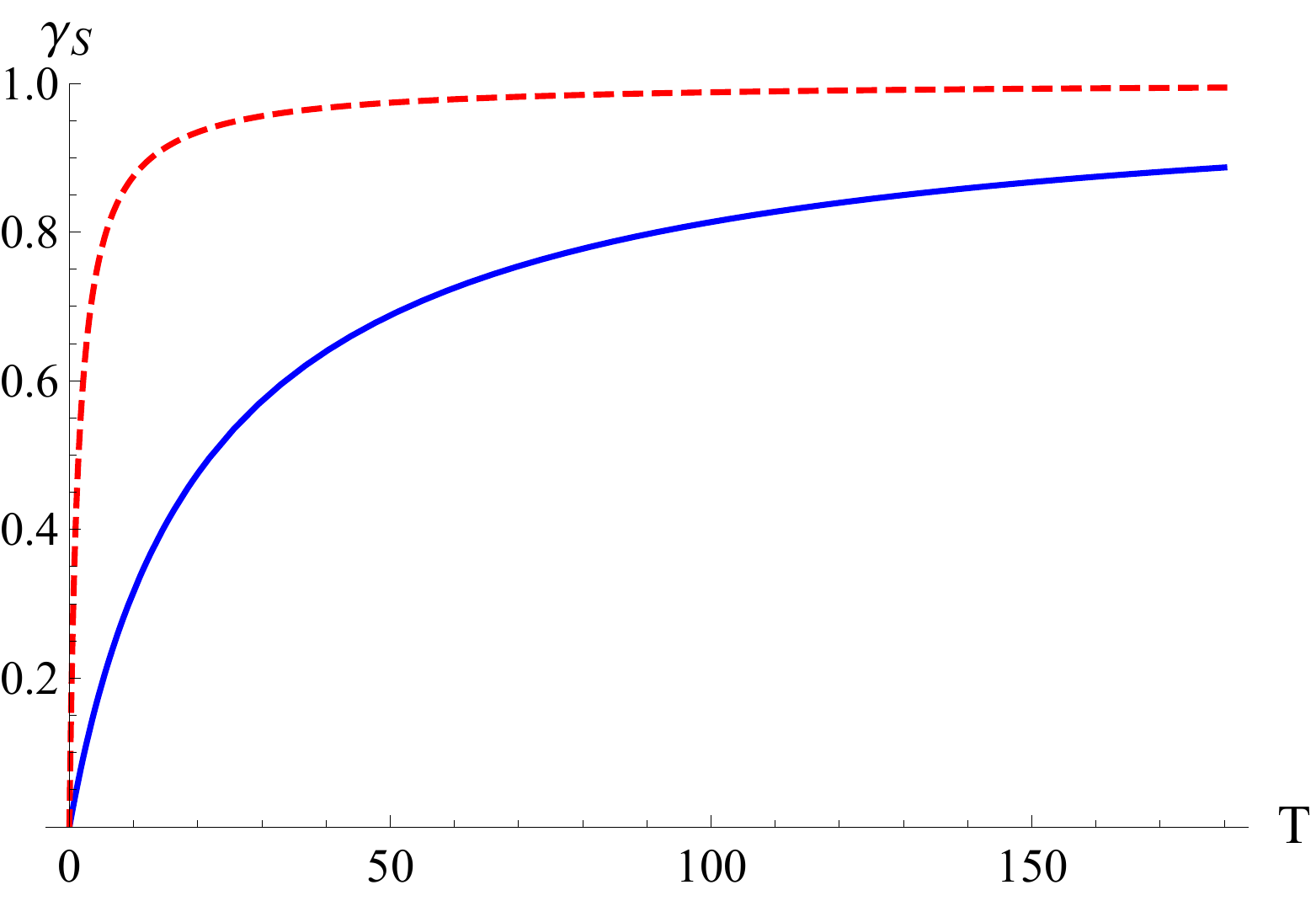}
	\caption{Normalized  spin relaxation rate as a function of temperature, eq.~\ref{gMnInt}. For both curves $l_D / a = 3$, $|\mathcal{B}_S - \mathcal{B}_J|/\gamma_J = 0.1$. Red dashed curve $T_1 = T_2 = 1$~K, blue curve $T_1 = 10$~K, $T_2 = 30$~K.}
	\label{figCalc}
\end{figure}

It is important to stress the difference and similarity between paramagnetic samples (with $Tc \rightarrow 0$) and ferromagnetic samples in the paramagnetic phase. The spin relaxation rate at the  saturation (see Eqs.\ref{gMnSat1}, \ref{gMnSat2}), is determined by the parameters $ M_0 $, $ D $, $ \gamma_J $ which monotonically vary with a change of the concentration delocalized carriers and magnetic centers. Whereas the saturation temperatures $ T_1 $, $ T_2 $ are defined by parameters determined by collective effects, which are critically sensitive to a change in the concentration of delocalized carriers and magnetic centers. This feature is clearly seen in GaMnAs. Thus, for a sample with manganese consentration $ x = 0.8 \% $ demonstrating paramagnetic behavior at temperatures above liquid helium, the spin relaxation saturation regime is achieved already at very low temperatures (see blue squares in right panel of Fig.~\ref{fig}). Whereas, for a sample with $ x = 1 \% $, which has an insignificant difference in concentration with the paramagnetic one, but has a pronounced ferromagnetic transition at $ T_C = 35$~K, a linear increase in the spin relaxation rate with subsequent saturation is observed (see red circles in right panel of Fig.~\ref{fig}). At the same time, the saturation values for these samples with quite similar concentrations differ insignificantly. 
In Fig.~\ref{figCalc} we present the model simulation of spin relaxation rate in paramagnetic phase for two samples with different Curie temperatures and similar $l_D / a = 3$, $|\mathcal{B}_S - \mathcal{B}_J|/\gamma_J = 0.1$. In this simulation we used the simpliest form of correlator (\ref{corrEx}). With this correlator the temperature dependence of $\gamma$ saturates slower than in the experiment but nevertheless demonstrate qualitative agreement with the  experiment in GaMnAs. To get better fit of the experimental data one needs to use a exact nonlinear correlator.

\section{Conclusion}
In conclusion we developed theory of magnetic impurity spin relaxation in DMS where carriers mediate ferromagnetism and simultaneously provide their spin relaxation. The theory assumes magnetic correlations therefore it can be applied for materials with high concentration of magnetic impurities. The theoretical model is compared with data on manganese transversal spin relaxation time measured by Raman spin-flip scattering in GaMnAs. It is found that for ferromagnetic sample in paramagnetic phase the spin relaxation rate saturates with temperature increase and tends to temperature independent value measured in paramagnetic sample with slightly lower manganese consentration.

\section*{ACKNOWLEDGMENTS}

This work was supported by Russian Science Foundation project No.~18-72-10111 (theory), project No.~18-12-00352 (experiment). Russian Foundation for Basic Research project  No.~19-02-00283 (comparison theory and experiment).


%

\end{document}